\documentclass[12pt]{article}
\usepackage{epsfig}

\newcommand{\ket}[1]{| #1 \rangle}

\newcommand{\braket}[2]{\langle #1 | #2 \rangle}

\def\gtap{\raisebox{-.55ex}{\rlap{$\sim$}} \raisebox{.4ex}{$>$}}
\def\gsim{\mathrel{\gtap}}

\def\e{\mbox{e}}
\def\ch{\mbox{ch}}
\def\sh{\mbox{sh}}
\def\th{\mbox{th}}
\def\tg{\mbox{tg}}

\begin{document}
\title{False vacuum decay in de Sitter space-time}
\author{
   V.~A.~Rubakov, S.~M.~Sibiryakov\\
    {\small \em Institute for Nuclear Research of the
Russian Aademy of Sciences,}\\
  {\small \em 117312, Moscow, 60th October Anniversary Prospect, 7a}\\
  }
\date{}
\maketitle
\begin{abstract}
We suggest a technique that explicitly accounts for the structure of
an initial state of quantum field in the semiclassical calculations
of path integral in curved space-time, and consider decay of
metastable state (conformal vacuum of scalar particles above false
classical vacuum) in background de Sitter space-time as an example.
Making use of this technique, we justify the Coleman--De Luccia
approach to the calculation of the decay probability. We propose an
interpretation of the Hawking--Moss instanton as a limiting case of
constrained instantons. We find that an inverse process of the transition
from true vacuum to false one is allowed
in de Sitter space-time, and calculate the corresponding probability.
\end{abstract}

\vskip .5 in

\section{Introduction}
The description of false vacuum decay in quantum field theory is
of considerable interest, especially in view of the developments of
inflationary theories and  models that explain the baryon
asymmetry of the Universe by invoking first order phase transitions.
It was established some time ago that the process of tunneling from
the false vacuum to the true one is semiclassical in weakly coupled
theories, and the adequate technique was developed for the calculation
of the decay probability in scalar field theories in Minkowski
space-time
\cite{Voloshin,Coleman,Col-Calan}. 
It was shown that the decay proceeds through the materialization of
bubbles of the new phase (true vacuum) in the metastable one (false
vacuum), and that the exponent in the semiclassical
expression for the probability of the formation of a bubble
 per unit time per unit volume,
\[
\Gamma = A\e^{-B}
\]
is determined by a solution to the classical field equations in
Euclidean space-time with the boundary conditions that require that
the field tends to the false vacuum as the Eucliden time
$\tau$ approaches $\pm \infty$ . 
The exponent $B$ is equal to the Euclidean action calculated on the
non-trivial (not identically equal to the false vacuum) solution of
this boundary problem, minus the action on the trivial solution. If
the boundary problem has several non-trivial solutions, then
the value of $B$ is determined by the least action one.
It was shown that for any scalar potential there exists an
$O(4)$-symmetric solution (bounce) \cite{Col-Calan}, 
and that it has the smallest action among non-trivial solutions
\cite{CGM}. The analytical continuation of this solution to pure imaginary
$\tau$, $\tau = it$, 
gives the configuration of the bubble after its materialization
as the function of the physical time
$t$. 
These results were obtained by making use of the path integral
technique. It is worth noting (although in this paper 
we will consider leading
semiclassical exponents only) that this technique allows one to
calculate pre-exponential factor
$A$ as well \cite{Col-Calan}. 

The problem of accounting for gravitational effects in the formation and
subsequent evolution of bubbles was studied by Coleman and De Luccia
\cite{CdL}. 
They proposed, by analogy to the case of flat space-time, that the
probability of the formation of a bubble is determined by 
an $O(4)$-symmetric solution to the system of equations consisting of
the
Klein--Gordon--Fock equation and Einsten equations for the scalar
field and Riemannian metrics with Euclidean signature. 
The exponent $B$ is again equal to the difference between the actions
of non-trivial and trivial solutions of this system (the trivial
solution is such that the scalar field is identically equal to its
false vacuum value), where the action contains now two terms: the
action of the scalar field in curved space-time and the action of the
gravitational field. 
It has been shown
\cite{Guth-W}, that if the scalar potential is non-negative (this
case is most interesting for inflationary theories), then the metrics
of this solution describes the space with the topology of a four-sphere
$S^4$.  
The analytical continuation of this solution to the space-time with
Minkowski signature gives the evolution of the bubble in real
space-time; the space-time in distant past is described by the de
Sitter metrics.

We think that this appealing and simple approach needs further
justification. In the first place, it is based solely on the analogy
to field theory in flat space-time. This analogy may in principle not
work 
 for space-times of large curvature. In particular, for large
class of scalar potentials, the only
$O(4)$-symmetric Euclidean solution is the Hawking--Moss instanton 
\cite{HawkMoss}, which is a four-sphere with constant 
scalar field. This instanton differs considerably from the bounce of
Minkowskian theory, and its naive analytical continuation to
space-time with Minkowskian signature would  correspond to
homogeneous scalar field rather than to the bubble of the true vacuum
in the false one.

Moreover, the very notion of the false vacuum in curved space-time
needs qualification. Besides specifying the classical vacuum --- the
average value of the scalar field --- one has also to specify the state
of quantum fluctuations of the field. Indeed, the vacuum state is not
uniquely defined in curved space-time even in the theory of free
quantum fields, as it depends on the choice of modes in which the
field operators are decomposed
(see, e.g., refs. \cite{Grib,Birel}). 
In this regard, it is of interest to establish what choice of quantum
vacuum leads to Coleman--De Luccia and Hawking--Moss prescriptions.
\footnote{It is likely that different choices of quantum states
above a given classical false vacuum will give rise to considerably
different decay probabilities. Indeed, there often emerges
the notion of the effective temperature that characterizes one or
another vacuum already in the theory of free fields (see, e.g., ref.
\cite{Birel}). On the other hand, the exponent for the decay of the
metastable state strongly depends on temperature in Minkowski
space-time \cite{Linde}.}

In a  certain range of the parameters of the theory, part of
these problems may be analyzed by making use of the stochastic approach 
\cite{Starob}. Within this approach, it has been demonstrated
\cite{Starob,GL} that the Hawking--Moss instanton action indeed
coincides with the exponent for the probability of the decay of the
classical false vacuum, if the state of quantum fluctuations above
this vacuum is the conformal vacuum of the scalar field. However, the
role of the Hawking--Moss solution itself remains unclear within the
stochastic approach.

A natural step in clarifying the above points is to study  the
false vacuum decay in the limiting case when the characteristic scale
of variation of the scalar potential during the tunneling process 
is small compared to the potential itself. In this case one can
neglect the changes of metrics in the process of the bubble formation,
so
one arrives at field theory in curved background space-time, namely, in the
de Sitter background. An obvious advantage of this approximation is
that one does not have to deal with quantum gravity effects, i.e., the
problem may be analysed in terms of ordinary field theory, albeit in
curved space-time.  We will restrict our analysis to the choice of the
quantum state above the classical false vacuum as the conformal vacuum
of scalar particles; this state appears naturally in inflationary
theories. Under these assumptions we will show that the decay of the
metastable state is indeed described by the Coleman--De Luccia
instanton (in models where it exists), give the interpretation of the
Hawking--Moss instanton as the limiting case of constrained instantons,
and discuss solutions that do not have analogs in flat space-time and
describe tunneling from the true vacuum to the false one.

\section{Preliminaries}
\begin{figure}[htb]
	\begin{center}
	\epsfig{file=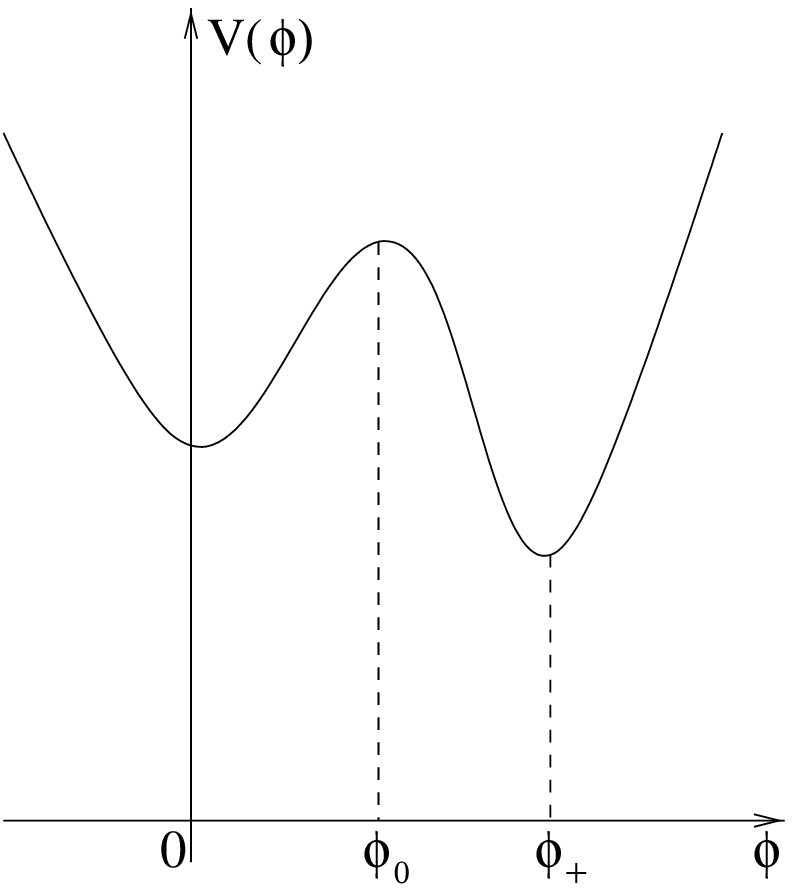, width=5cm}
	\caption{}
	\label{1f}
	\end{center}
\end{figure}

We consider the theory of the scalar field $\phi$ with the potential
shown in fig. \ref{1f}.
The states $\phi = 0$ and $\phi = \phi_+$ correspond to the false and
true vacua, respectively; they are separated by the local maximum of
the scalar potential, 
$V(\phi_0)$.
Let us first discuss the following question: does it make sense at all
to consider gravitational effects on tunneling in the
background
metrics approximation ? Let us introduce the mass scale
$M$ that measures the variation of the potential in the interval
between the two vacua.
The gravitational field is characterized by the Hubble parameter $H$. 
The gravitational effects are important for the formation of the
bubble when the ratio $\frac{H}{M}$ is of order or larger than 1,
while the approximation of background de Sitter metrics is valid at
\begin{equation}
   \frac{M^4}{V(\phi_{+})} \ll 1\,.
\label{ineq*}
\end{equation}
Making use of the Einstein equation,
\[
	H^2 = \frac{8\pi}{3M_{Pl}^2}V(\phi_+)
\]     
one finds
\[
	\frac{H}{M} = \frac{1}{MM_{Pl}}\sqrt{\frac{8\pi V(\phi_+)}{3}}
\]
This ratio can easily be of order one, if
$\frac{V(\phi_+)^{\frac{1}{4}}}{M_{Pl}} \gsim 
\frac{M}{V(\phi_+)^{\frac{1}{4}}}$. 
At the same time, the inequality
(\ref{ineq*}) may well be satisfied, and the energy density
$V(\phi_+)$ may be small compared to the Planck value.
Hence, the approximation of background metrics is legitimate in some
region of parameters, and in the same region the gravity effects are not
negligible.

\begin{figure}[htb]
	\begin{center}
	\epsfig{file=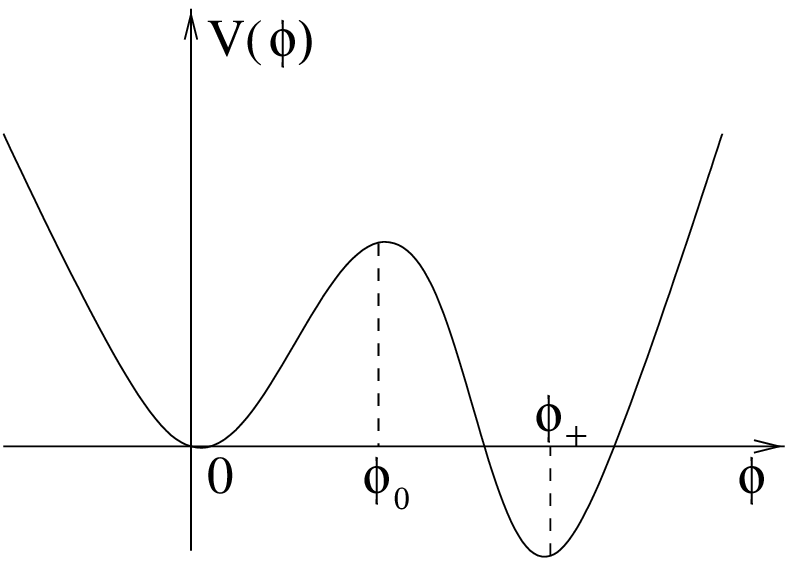, width=5cm}
	\caption{}
	\label{f2}
	\end{center}
\end{figure}

In what follows we will consider the scalar field in the background
metrics, so we may freely change the origin of its energy. It is
convenient to set $V(0)=0$, as shown in
fig. \ref{f2}.

The choice of the state of  quantum fluctuations above the
classical false vacuum is related to the choice of reference
frame. We will
use the frame with flat surfaces of constant time, in which the de
Sitter metrics takes the following form,
\begin{equation}
	dl^2 = \frac{1}{H^2\eta^2}(d\eta^2 - dx_idx_i)
\label{1.2}
\end{equation}
The conformal time parameter $\eta$ changes from $-\infty$ 
to $0$; it is related to the sinchronous time 
$t$ as $\eta = -\frac{1}{H} e^{-Ht}$. This coordinate system
covers only one half of the full de Sitter space (see, e.g., ref.
\cite{Hawk-Ellis}), but it is this part that is of interest to
inflationary models. The action of the model in these coordinates is
\begin{equation}
	S = \int\frac{d^3xd\eta}{H^4\eta^4}\left[\frac{H^2\eta^2}{2}
	\left(\left(\frac{\partial\phi}{\partial\eta}\right)^2 - 
	\left(\frac{\partial\phi}{\partial x^i}\right)^2\right) -
	V(\phi) \right]
\label{1.3}
\end{equation} 
Let us decompose the quantum perturbations about the false vacuum
in scalar modes which are natural for this frame
\cite{Birel},
\[
	\hat{\phi}({\bf x},\eta) =
    \int\frac{d^3k}{(2\pi)^{\frac{3}{2}}} H \eta \left(a_k\chi_k(\eta)
    + a^+_{-k} \chi^*_k(\eta)\right) e^{i{\bf k}{\bf x}} 
\]
\[
	\chi_k(\eta) = \frac{(\pi\eta)^{\frac{1}{2}}}{2}
    H_{\nu}^{(2)}(k\eta),~~~~\nu^2 = \frac{9}{4} -
    \frac{m^2}{H^2},~~~~m^2 = V''(0),~~~~k=|{\bf k}|
\]
Here $H_\nu$ are the Hankel functions.
The conformal vacuum above the classical false one is 
a state that is annihilated by all operators
$a_k$: $a_k\ket{0} = 0$.  Though this definition makes sense in
perturbation theory only, it will be sufficient for our purposes.
Our choice of frame, and related choice of quantum vacuum is due to
several reasons:

\begin{enumerate}
\item
This frame and this definition of vacuum appear naturally in
inflationary theories (see., e.g., ref.
\cite{infl_cosm}).
\item
This vacuum is invariant under spatial translations,
$x \to x+a$, and scale transformations,
$\eta \to \alpha\eta$, $x \to \alpha x$,
which are the global isometries of the half of the de Sitter space.
\item
The modes $\chi_k$ reduce to negative frequency plane waves at large
negative  $\eta$,
\begin{equation}
	\chi_k\to\frac{1}{(2k)^{\frac{1}{2}}}
	e^{-ik\eta},~~~~\eta\to\infty
\label{1.7}
\end{equation}
So, one expects that at large negative times the theory in coordinates
$(\eta,{\bf x})$ is similar to the theory in flat space-time, and this is
indeed the case (see below).
\end{enumerate}

The  false vacuum state, as defined  above, is metastable. 
The problem is to find the leading semiclassical exponent for the
probability of its decay.

\section{Classical problem and boundary conditions}

In this Section we show that in the semiclassical regime, the exponent
for the probability of the false vacuum decay is indeed determined by
a certain classical solution to the scalar field equation, and obtain
the boundary conditions that are to be imposed on the solution.

Let us introduce the following notations:
$\ket{n,\eta}$ is a state $n$ at time $\eta$
(we work in the Heisenberg representation); in particular,
$\ket{0,\eta}$ is the state of the false vacuum at  time
$\eta$, $\ket{\phi({\bf x}),\eta}$ 
is the common eigenstate of the field operators taken
at time $\eta$.
To calculate the probability of the false vacuum decay, we will need
the
matrix element
$\braket{\phi_f({\bf x},\eta_f)}{0,\eta_i}$, where $\eta_f$ and $\eta_i$
are the final and initial moments of time, respectively.
This matrix element can be written in terms of path integral,
\begin{equation}
        \braket{\phi_f({\bf x},\eta_f)}{0,\eta_i} =
        \int D[\phi_i({\bf x})]\braket{\phi_i({\bf x}),\eta_i}{0,\eta_i}
        \int D[\phi({\bf x},\eta)] e^{iS[\phi]}
\label{2.1}
\end{equation}
Here $S[\phi]$  is the action in the interval
$(\eta_i,\eta_f)$ evaluated on the field configuration
$\phi ({\bf x},\eta)$,
obeying the boundary conditions
\[
        \phi({\bf x},\eta_i) = \phi_i({\bf x}),~~\phi({\bf x},\eta_f)
        = \phi_f({\bf x})
\]
The outer integral is over the initial configurations of the
field. Two comments are in order:
\begin{enumerate}
\item
Time interval $(\eta_i, \eta_f)$ is taken to be much larger than any
time scale of the theory. We are ultimately interested in the limit
$\eta_i \to - \infty$.
\item
Until now the variables ${\bf x}$ and $\eta$  were
real coordinates in the de Sitter space-time. However, one can consider
the space of complex $({\bf x},\eta)$ 
and analytical continuations of the metrics and field into this space.
The four-dimensional hypersurface over which the action in eq.
(\ref{2.1}) is integrated, may be deformed arbitrarily in this space, 
provided that its boundaries, the three-dimensional initial
and
final surfaces, remain intact.
\end{enumerate}

The decay width of the false vacuum takes the following form,
\begin{equation}
        \Gamma = \int D[\phi_f({\bf x})]~
	|\braket{\phi_f,\eta_f}{0,\eta_i}|^2 \, ,
\label{2.2}
\end{equation}
where the integration runs over the configurations
$\phi_f({\bf x})$
that are close to the true vacuum (the meaning of the latter
requirement will become clear later on).
Upon substituting eq.
(\ref{2.1}) into eq. (\ref{2.2}) one gets
\begin{eqnarray}
        \Gamma &=& \int D[\phi_f({\bf x})]
        \int D[\phi_{1i}({\bf x})]  D[\phi^*_{2i}({\bf x})]
        \braket{\phi_{1i},\eta_i}{0,\eta_i}
        \braket{\phi_{2i},\eta_i}{0,\eta_i}^* \nonumber \\
    &\times&    \int D[\phi_1({\bf x},\eta)] D[\phi_2({\bf x},\eta)]
        e^{i(S[\phi_1] - S^*[\phi_2])}
\label{2.3}
\end{eqnarray}
Our immediate purpose is to evaluate the path integrals in eq. 
(\ref{2.3}) by the saddle point technique (which of course assumes
that the problem is semiclassical). Let us first integrate over
$D[\phi_1]D[\phi_2]$.  The saddle point of the integrand is a solution
of the classical field equations. The corresponding classical solutions
$\phi_1({\bf x},\eta)$ and $\phi_2({\bf x},\eta)$
obey the following conditions,
\begin{equation}
        \phi_1({\bf x},\eta_i) = \phi_{1i}({\bf x}),~~
        \phi_2({\bf x},\eta_i) = \phi_{2i}({\bf x}),~~
        \phi_1({\bf x},\eta_f) = \phi_2({\bf x},\eta_f) =
        \phi_f({\bf x})
\label{2.5}
\end{equation}
To integrate over $D[\phi_{1i}]D[\phi_{2i}]$,
one needs explicit expressions for the matrix elements of the type
$\braket{\phi_i({\bf x}),\eta_i}{0,\eta_i}$.
Their calculation parallels the case of flat space-time (this is due
to the fact that the modes
$\chi_k$ tend to plane waves as $\eta \to - \infty$).
Since the relevant matrix elements are large only for field
configurations close to the false vacuum, one makes use of
perturbation theory. Let us introduce the momenta conjugate to the
field operators,
\begin{eqnarray}
        \hat{\pi}({\bf x},\eta) &=& \frac{1}{H^2\eta^2}
                 \frac{\partial\hat{\phi}}{\partial\eta} \nonumber \\
          &=& \int
        \frac{d^3k}{(2\pi)^{\frac{3}{2}}} \frac{1}{H^2\eta^2}
        (a_k\chi_k(\eta) + a^+_{-k}\chi^*_k(\eta)) e^{i{\bf k}{\bf x}}
        \nonumber \\
        &+& \int
        \frac{d^3k}{(2\pi)^{\frac{3}{2}}} \frac{1}{H\eta}
        (a_k\dot{\chi}_k(\eta) + a^+_{-k}\dot{\chi}^*_k(\eta))
        e^{i{\bf k}{\bf x}}
\label{2.6}
\end{eqnarray}
Note that the first term in eq. (\ref{2.6}) is negligible at large negative
$\eta$. By introducing the spatial Fourier components, 
\[
        \hat{\phi}({\bf k},\eta_i) = H\eta_i (a_k\chi_k(\eta_i) +
        a^+_{-k}\chi^*_k(\eta_i)),
        ~~~\hat{\pi}({\bf k},\eta_i) = \frac{1}{H\eta} (a_k\dot{\chi}_k(\eta_i) +
        a^+_{-k}\dot{\chi}^*_k(\eta_i))
\]
one obtains the standard commutational relations
\[
        [\hat{\pi}(-{\bf k},\eta_i), \hat{\phi}({\bf q}, \eta_i)] =
        -i\delta({\bf k}-{\bf q})
\]
Hence, the operators $\hat{\phi}({\bf k},\eta_i)$ and
$\hat{\pi}({\bf k},\eta_i)$ may be realized as multiplication by
$\phi({\bf k})$ and variational derivation,
$\frac{-i\delta}{\delta\phi(-{\bf k})}$, respectively.
In this way one arrives at the coordinate representation.
The state 
$\ket{0,\eta_i}$ in this representation is the functional
$F[\phi({\bf k})] =
\braket{\phi({\bf k}),\eta_i}{0,\eta_i}$. One obtains the equation for
this functional by exressing the operators
 $a_k$ through $\hat{\phi}({\bf k},\eta_i)$ and 
$\hat{\pi}({\bf k},\eta_i)$  and recalling that
$a_k$ annihilate the vacuum,
\begin{equation}
        \left(\frac{\delta}{\delta\phi(-{\bf k})} +
        \frac{k}{(H\eta_i)^2} \phi({\bf k})\right) F = 0
\label{2.9}
\end{equation}
(to obtain eq. (\ref{2.9}) one makes use of the asymptotics (\ref{1.7})). 
Hence,
\begin{equation}
        F = F_0  \exp\left[ {-\int
        \frac{k\phi({\bf k})\phi(-{\bf k})}{2(H\eta_i)^2} d^3k}\right]
\label{2.10}
\end{equation}
The normalization constant $F_0$ will be inessential for our purposes.

Let us substitute eq. (\ref{2.10}) into eq. (\ref{2.3}) and integrate over
$D[\phi_{1i}]$. The saddle point of the integrand is determined by the
condition
\[
        \frac{\delta}{\delta\phi_{1i}(-{\bf k})}
        \left(-\int \frac{k\phi_{1i}({\bf k})\phi_{1i}(-
        {\bf k})}{2(H\eta_i)^2} d^3k + iS[\phi_1]\right) = 0
\]
which implies that
\[
        \frac{k}{(H\eta_i)^2}\phi_{1i}({\bf k}) +
        i\frac{1}{(H\eta_i)^2} \dot{\phi_1}({\bf k},\eta_i) = 0
\]
In this way we obtain the positive frequency condition at
$\eta \to -\infty$ that has to be imposed on the classical solution
$\phi_1$.  The same argument leads to the 
positive frequency condition on
$\phi_2$ at large negative $\eta$. The boundary conditions for
$\phi_1$ and $\phi_2$ at $\eta_f$ are also the same 
(see eq. (\ref{2.5})); furthermore,   $\phi_1$ and $\phi_2$
are solutions to the same classical field equation. So,
$\phi_1$ and $\phi_2$ are actually one and the same function
\footnote{Indeed, a general solution to the second order equation
contains two arbitrary complex functions of three variables. The
positive frequency condition kills one of these degrees of
freedom. The second function is determined by the boundary condition
imposed at the final time.}
which we denote merely by
$\phi$. 
Thus, the saddle point integration over the intermediate and initial
values of the field gives
\begin{equation}
        \Gamma \propto \int D[\phi_f] e^{i(S[\phi] - S^*[\phi])}
\label{2.15}
\end{equation}
where $\phi({\bf x},\eta)$ is (in general, complex) solution of the
classical scalar field equation in the de Sitter background metrics, 
obeying the boundary conditions

\[
        \phi({\bf x},\eta_f) = \phi_f({\bf x})
\]
\begin{equation}
        \phi ~~~
\mbox{contains only positive frequency waves  at}~~ 
	 \eta \to -\infty
\label{2.15b}
\end{equation}
As we are going to discuss the leading semiclassical exponent only,
the pre-exponential factor is not shown in eq. (\ref{2.15}).
Note that the boundary condition (\ref{2.15b}), in similarity to
Minkowski background, allows one not only to deform the integration
contour but also to shift the boundary ``point''
$\eta_i = -\infty$
in upper complex half-plane of the variable
$\eta$. 
Hereafter the integration contour (or simply contour) means the
hypersurface used in the calculation of the action
in eq. (\ref{2.15}).

There still 
remains one  integration in eq. (\ref{2.15}). We find the
corresponding saddle point equation: 
\[
        \dot{\phi}({\bf x},\eta_f) - \dot{\phi}^* ({\bf x},\eta_f) = 0
\]
which, together with reality of $\phi_f({\bf x}$) gives the following
condition, 
\begin{equation}
        \phi({\bf x},\eta)~~~\mbox{is real at}~ \eta > \eta_f
\label{2.17}
\end{equation}
When the solution
$\phi(\eta)$ (we temporarily ignore the dependence of 
$\phi$ on spatial coordinates) is continued along the real axis
from the domain
$\eta >\eta_f$ to the region $\eta
<\eta_f$, it will remain real. 
This property may seem to contradict the condition
(\ref{2.15b}).  However, this observaion only implies that a
solution with required properties does not exist on the real axis, so
one has to search for such a solution on the deformed contour that
avoids singular points of the function
$\phi$ (cf. ref. \cite{Rub}).

\begin{figure}[htb]
        \begin{center}
        \epsfig{file=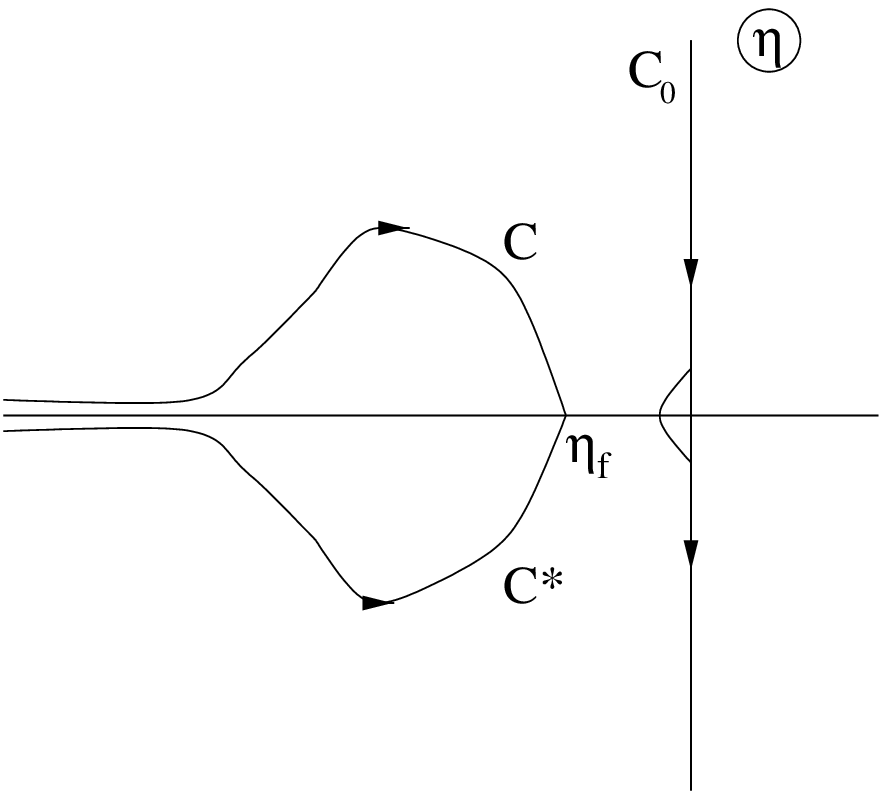, width=6cm}
        \caption{}
	\label{f3}
        \end{center}
\end{figure}

The values of $\phi$
in upper the and lower half-planes are related by
$\phi(\eta^*) = \phi^*(\eta)$.
Taking this property into account, we rewrite the exponent in
eq. (\ref{2.15}) as follows,
\[
        i(S[\phi] - S^*[\phi]) = i(S_C[\phi] - S_{C^*}[\phi])
        = iS_{C+C^*}[\phi]
\]
Here $S_C$, $S_{C^*}$ and $S_{C+C^*}$ stand for the values of the
action
at the contours 
$C$, its conjugate contour $C^*$ and their sum
(see fig. \ref{f3}). Now we can relax the condition that the contours
$C$ and $C^*$ meet at the point
$\eta_f$ and deform the contour $C+C^*$ into a contour
$C_0$ that consists of the imaginary axis and infinitesimal
semi-circle around the singular point of the metrics,
$\eta = 0$.

It is convenient to introduce a new coordinate $\zeta$  such that
$\eta = i\zeta$, and rewrite the action in the follwing form,
\begin{equation}
        iS_{C_0}[\phi]
    =    \int d^3x \int\limits_{-\infty}^{\infty ~\prime} d\zeta
        \frac{1}{H^4\zeta^4} \left[\frac{H^2\zeta^2}{2}
        \left(\left(\frac{\partial\phi}{\partial\zeta}\right)^2 +
         \left(\frac{\partial\phi}{\partial x^i}\right)^2\right) -
        V(\phi)\right] \equiv -S_{\zeta}[\phi]
\label{2.19}
\end{equation}
(Primed integral denotes the integration near $\zeta=0$ along the
contour $C_0$.)

Let us summarize the outcome of the analysis made in this Section.

\begin{enumerate}
\item
The leading semiclassical exponent of the false vacuum decay may be
written as follows,
\begin{equation}
       \Gamma \propto e^{-S_{\zeta}[\phi]}
\label{2.20}
\end{equation}
\item
The function $\phi$ in eq. (\ref{2.20}) is a classical solution to the
field equation at the contour
$C_0$.
\item
The
analytical continuation of $\phi$ 
to the real values of $\eta$ should be real.
\item
$\phi(\zeta)$ tends to zero as
$\zeta \to\infty $
(the latter condition is a consequence of eq.
(\ref{2.15b})).  
\end{enumerate}

To deform the integration contour (four-dimensional hypersurface
in the space of complex
${\bf x}$ and $\eta$)  further, it is useful
to discuss the analytical continuations of the de Sitter metrics.

\section{Analytical continuation of metrics}

In the previous Section we introduced the coordinate
$\zeta$ related to
$\eta$ by $\eta = i\zeta$. 
Let us first analyze the metrics of space with coordinates
$({\bf x},\zeta)$ 
at $0<\zeta<\infty$. 
 It is clear from eq.(\ref{1.2}) that the line element in this space is
\begin{equation}
       dl^2 = \frac{d\zeta^2 + dx_i^2}{H^2\zeta^2}
\label{3.1}
\end{equation}
that is, the metrics has Euclidean signature. This is the space of
constant negative curvature, as is clear directly from the
corresponding property of the de Sitter space-time
\cite{Birel}.  However, it can be shown that this space is
geodesically complete, unlike the half of the de Sitter space that has
been considered above.

Let us introduce spherical coordinates in three-dimensional space
with coordinates
$x_i$; then eq. (\ref{3.1}) is written as follows,
\begin{equation}
       dl^2 = \frac{d\zeta^2 + d\rho^2 + (d\Omega_2)^2}{H^2\zeta^2}
\label{3.2}
\end{equation}
where $\rho = \sqrt{x_1^2+x_2^2+x_3^2}~$, and $~(d\Omega_2)^2 =
d\theta^2+\sin{\theta}^2 d\varphi^2,~~0\le\theta \le \pi,~~0\le\varphi
<2\pi$ is the line element on unit two-dimensional sphere.
Until now, the coordinates
$x,\eta,\zeta,\rho$  had the dimension of length. However, it is
convenient to make these coordinates
dimensionless, i.e., measure them in units of
the inverse Hubble parameter. The line element is not changed after
this redefinition, while the expression for the action changes only
slightly, 
\begin{equation}
        S_{\zeta} = -\int d^3x \int\limits_{-\infty}^{\infty~\prime} d\zeta
        \frac{1}{\zeta^4} \left[\frac{\zeta^2}{2H^2}
        \left(\left(\frac{\partial\phi}{\partial\zeta}\right)^2 +
        \left(\frac{\partial\phi}{\partial x^i}\right)^2\right) -
        \frac{V(\phi)}{H^4}\right]
\label{3.3}
\end{equation}

\begin{figure}[htb]
        \begin{center}
        \epsfig{file=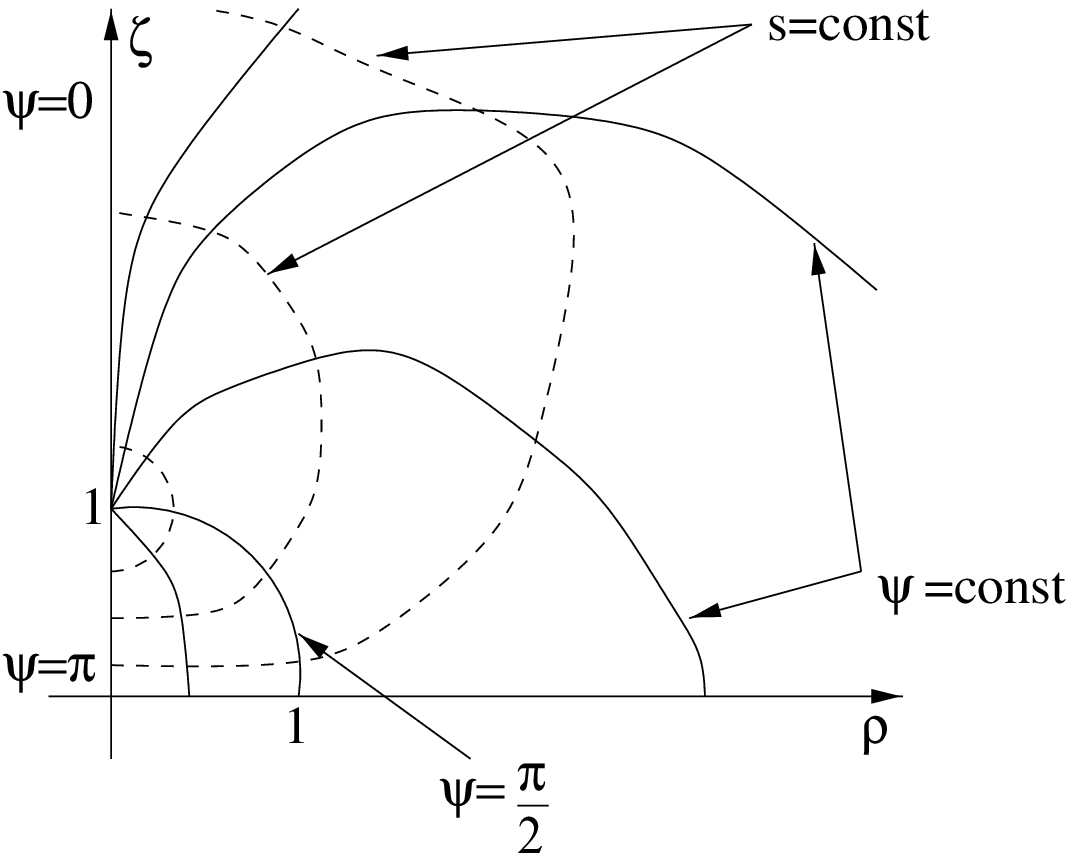, width=8cm}
        \caption{}
	\label{f4}
        \end{center}
\end{figure}

Let us introduce, in the space under discussion, the coordinates that
are reminiscent of the usual spherical coordinates. Namely, let us
choose a point (call it the origin) with coordinates
$\rho=0,\zeta=\zeta_0$, and consider all possible geodesics that start
from this point. The distance along these geodesics will be our radial
coordinate $s$.
Explicitly,
\[
       s = \int\limits_{(0,\zeta_0)}^{(\rho,\zeta)} dl =
       \int\limits_{(0,\zeta_0)}^{(\rho,\zeta)} \frac{d\zeta'}{\zeta'}
       \sqrt{1+\left(\frac{d\rho }{d\zeta' }\right)^2}
\]
The integration here is performed along the geodesics that connects the
points 
$(\rho ,\zeta )$ and $(0,\zeta )$, the derivative
$\frac{d\rho }{d\zeta' }$ is understood  correspondingly.
Note that the  value of 
$\zeta _0$ is actually unimportant, as its change is 
equivalent to the change of scale. In what follows we set it equal to
1 for definiteness. Straigtforward but tedious calculations lead to
the following formulae that express the coordinates
 $\rho $  and $\zeta $
through $s$ and an angular variable $\psi $  characterizing the
direction of the relevant geodesics,
\begin{equation}
      \zeta =\frac{1}{\ch{s}-\cos{\psi }~\sh{s}},~~~
      \rho = \frac{\sin{\psi }~\sh{s}}{\ch{s}-\cos{\psi  }~\sh{s}},~~
      0\le s<\infty,~~ 0\le\psi \le\pi
\label{3.5}
\end{equation}
Lines of constant $s$ and $\psi $ are shown in fig. \ref{f4}.
The line element in the new coordinates is
\begin{equation}
      dl^2 = \frac{1}{H^2} (ds^2 + \sh^2s~(d\psi ^2+\sin{\psi
      }^2(d\Omega_2)^2))
\label{3.6}
\end{equation}

Let us discuss the change of coordinates (\ref{3.5}) in more detail. 
Equation $\psi =0$ describes a ray $\rho =0$, $1 \le \zeta<\infty$
on the plane $(\rho ,\zeta)$  while equation
$\psi =\pi $ corresponds to the interval $\rho =0$, $0<\zeta \le
1$. The line $\psi 
=\frac{\pi}{2}$ is a part of a circle $\rho ^2+\zeta ^2=1$. 
If one sends $s$ to infinity keeping 
$\psi \ne 0$ fixed,  then
\begin{equation}
       \zeta \sim \frac{e^{-s}}{1-\cos{\psi }},~~\rho \to
       \frac{\sin{\psi }}{1-\cos{\psi }},~~s\to\infty
\label{3.7}
\end{equation}
Finally, the asymptotics $s\to\infty$, $\psi \sim e^{-2s}$ corresponds
to
$\rho \to \mbox{const}$, $\zeta \to\infty$.

Let us now consider $s$ as a complex variable
(while keeping $\psi $  real) and analyse the corresponding
analytical continuations of the metrics.
The following cases will be of interest for our purposes:
\begin{enumerate}
\item
$s=-i\pi +\hat{s}$ , where $\hat{s}$ is real. Upon subsituting this
relation into eq. (\ref{3.5}) 
we find that
 $\rho$  is expressed through $\hat{s}$ in the same way
as it has been expressed throgh
$s$, 
and the formula for
$\zeta $  changes by the sign only.
Hence, these values of $s$ describe the space with negative
$\zeta $; the structure of this space is completely analogous to the
case 
of positive $\zeta $.
\item
$s = -i\frac{\pi }{2}+s'$, where $s'$  is real.
In this case one has 
\begin{equation}
       \rho =\frac{\sin{\psi }~\ch{s'}} {\sh{s'}-\cos{\psi }~\ch{s'}},~~~
       \zeta =\frac{i}{\sh{s'}-\cos{\psi }~\ch{s'}}
\label{3.8}
\end{equation}
Under the restriction
$\cos{\psi }< \th{s'}$ 
these formulae describe space with
$\rho >0$ and pure imaginary $\zeta $ that changes from $0$ to
$+i\infty$.  This space is nothing but half of the de Sitter
space-time discussed in Sections 2 and 3 (the restriction imposed on 
$s'$ at given $\psi $  is of course related to the geodesic
incompleteness of this part of the de Sitter space).
Lines of constant $s'$, $\psi $ on the plane
$\eta =i\zeta$, $\rho $ have the form shown in 
fig. \ref{5f}.
\begin{figure}[htb]
        \begin{center}
        \epsfig{file=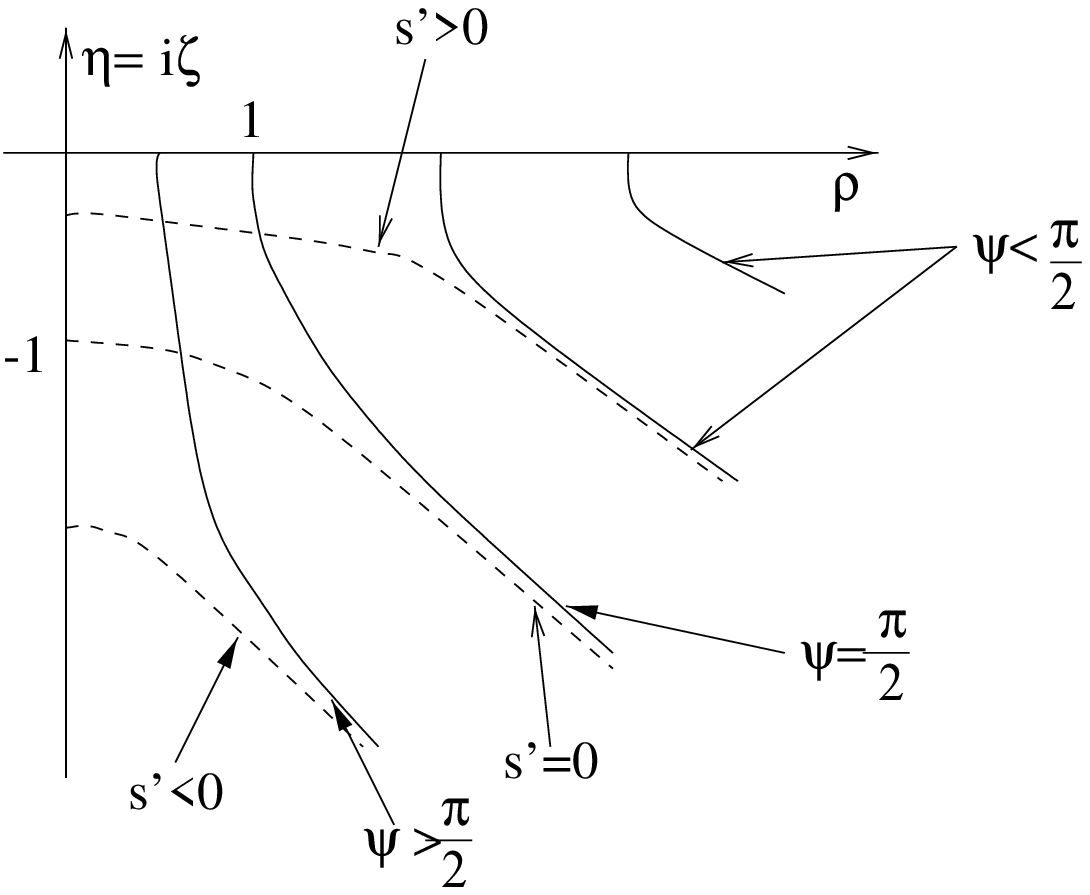, width=8cm}
        \caption{}
	\label{5f}
        \end{center}
\end{figure}
This space is characterized by the metrics
\begin{equation}
        dl^2 = \frac{1}{H^2} (ds'^2-\ch^2s'~(d\psi ^2+\sin^2\psi
        (d\Omega_2)^2))
\label{3.9}
\end{equation}
\item
$s=-i\sigma$, $\sigma $  
is real and belongs to $[0,\pi ]$. 
In terms of $\sigma$ the metrics (\ref{3.6}) is
\begin{equation}
       dl^2 = \frac{1}{H^2} (-d\sigma ^2- \sin^2\sigma~ (d\psi ^2 +
       \sin^2\psi  (d\Omega_2)^2))
\label{3.10}
\end{equation}
which is the metrics of four-sphere (up to sign).
\end{enumerate}

\section{Coleman--De Luccia solution}

In the approximation of background (non-dynamical) metrics, the
Coleman--De Luccia configuration is the solution to the scalar field
equation on four-sphere. In this Section we show that this
configuration is indeed a solution to the boundary problem formulated
in Section 3. 

\begin{figure}[htb]
        \begin{center}
        \epsfig{file=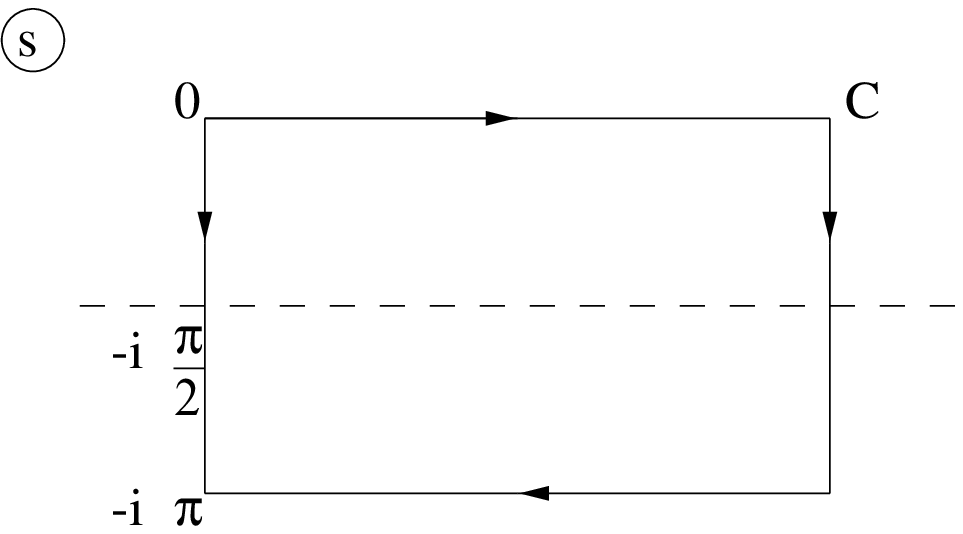, width=6cm}
        \caption{}
	\label{f6}
        \end{center}
\end{figure}

Let us rewrite the action integral in eq.  (\ref{3.3}) as an integral
over 
the four sphere. We begin by noting that eq.(\ref{3.3}) 
has the following general
form,
\begin{equation}
       S_\zeta = -\int\limits_{0}^{\infty} \rho ^2 d\rho
       \int\limits_{-\infty}^{\infty ~\prime} \frac{d\zeta }{\zeta ^4}
       f(\rho,\zeta)
\label{4.1}
\end{equation}
where $f(\rho ,\zeta )$ is an analytic function of two variables. Let
us split this integral into the sum of the three terms: the integrals
over positive and negative
$\zeta $ and the integral near the singular point
$\zeta =0$. Let us change variables in the first of these integrals, 
\[
       \int\limits_{0}^{\infty} \rho ^2 d\rho \int\limits_{0}^{\infty}
       \frac{d\zeta }{\zeta ^4} f(\rho ,\zeta ) =
       \int\limits_{0}^{\pi } d\psi \int\limits_{0}^{\infty}
       ds~\sh^3s~ \sin^2\psi~  \tilde{f}(\psi ,s),~~~ 
\]
where
\[
       \tilde{f}(\psi ,s) = f\left(\frac{\sin{\psi}~\sh{s}}
       {\ch{s}-\cos{\psi }~\sh{s}}, \frac{1}{\ch{s}-\cos{\psi
       }~\sh{s}}\right)
\]
The second term may be written in a similar way,
\[
       \int\limits_{0}^{\infty} \rho ^2 d\rho
       \int\limits_{-\infty}^{0} \frac{d\zeta }{\zeta ^4} f(\rho,\zeta ) =
       \int\limits_{0}^{\pi } d\psi \int\limits_{0}^{\infty}
       d\hat{s}~\sh^3\hat{s}~ \sin^2\psi~
       f\left(\frac{\sin{\psi}~\sh{\hat{s}}}
       {\ch{\hat{s}}-\cos{\psi }~\sh{\hat{s}}},
       -\frac{1}{\ch{\hat{s}}-\cos{\psi }~\sh{\hat{s}}}\right)
\]
This expression can be further rewritten as follows (cf. Section 4),
\[
       -\int\limits_{0}^{\pi } d\psi \int\limits_{-i\pi }^{-i\pi +\infty}
       ds~\sh^3s~ \sin^2\psi~ \tilde{f}(\psi ,s)
\]
Hence, the integration over positive and negative
$\zeta $ corresponds to the integration 
 of one and the same analytical function 
$\tilde{f}$ along the rays in the complex
$s$-plane,
$[0,\infty)$ and $(\infty-i\pi ,-i\pi ]$,  respectively.
To obtain the complete integral
(\ref{4.1}), one has to close the contour.
It is clear from eq.
(\ref{3.7}) that $\zeta $ is small when
$s$  is large, so an interval
$[s_1,s_1+i\pi ]$  at large real
$s_1$ corresponds to a path above the singularity 
$\zeta =0$ in the complex $\zeta$-plane, i.e., exactly the path
that is understood in the integral
(\ref{4.1}).  
Hence, the contour has to be closed at infinity, and the action
$S_\zeta $  is equal to the following expression,
\begin{equation}
       S_\zeta = -\int\limits_{0}^{\pi }d\psi \int\limits_{C} 
	ds~\sh^3s~ \sin^2\psi~
       \tilde{f}(\psi ,s)
\label{4.5}
\end{equation}
where the contour  $C$ is shown in fig. \ref{f6}.
This contour can be deformed into an interval $[0,\pi]$ on 
the imaginary axis.
The action nicely takes the form of the integral over the four-sphere,
\begin{equation}
       S_\zeta = -\int\limits_{0}^{\pi }d\psi \int\limits_{0}^{\pi }d\sigma
       \sin^3\sigma~ \sin^2\psi~ \tilde{f}(\psi ,-i\sigma )
\label{4.6}
\end{equation}
To complete the transition to the four-sphere, we have to reformulate
the field equations and boundary conditions imposed on the classical
solution (see the end of Section 3). The former step is easy: one has
just to find the explicit form of the function 
$\tilde{f}$ entering eq. (\ref{4.6}) that is obtained from
eq. (\ref{3.3}) after the change of variables,
\begin{equation}
       \tilde{f}(\psi ,-i\sigma ) = -\int\limits_{0}^{\pi}
       d\theta\sin{\theta} \int\limits_{0}^{2\pi }d\varphi
       \left(\frac{(\nabla_{S^4}\phi )^2}{2H^2} + \frac{V(\phi)}{H^4}\right)
\label{4.7}
\end{equation}
where $\nabla_{S^4}$ denotes the gradient on four-sphere.
(Note that 
the gradient and potential terms enter now with the same
sign, i.e., the scalar potential has got reversed, cf. refs.
\cite{Coleman,Col-Calan,CdL}.) By varying the action, one obtains the
standard equation
\begin{equation}
       \Delta_{S^4} \phi = \frac{V'(\phi )}{H^2}
\label{4.8}
\end{equation}
where $\Delta_{S^4}$ is the Laplace operator on the four-sphere.

The requirement of real-valuedness of $\phi$ at real $\eta$ (see
Section 3) is also easily formulated in terms of the solution on the
four-sphere. It is equivalent to the requirement that
$\phi $ is real at $ s= -i\frac{\pi }{2}+s' $ where
$s' $  is real (see Section 4).  The latter leads to the relation
\begin{equation}
       \left.\phi \right|_{s=z-i\frac{\pi}{2}} =
	\left.\phi ^*\right|_{s=z^*-i\frac{\pi }{2}}
\label{4.9}
\end{equation}
Let us make an important assumption that
the solution
$\phi $  is real on the sphere, i.e., at
$s= -i\sigma $. This assumption, together with eq. (\ref{4.9}), gives
the 
following condition:

C1.  For the solution $\phi $ be real at real
$\eta$ it is sufficient and necessary that $\phi $ is symmetric on the
sphere with respect to the ``equator''
$\sigma =\frac{\pi}{2}$.

The requirement that 
$\phi $  vanishes as
$\zeta $ tends to $\infty$ (see the end of Section 3)
cannot be formulated in terms of the four-sphere, so one has to
leave this condition as

C2. Analytical continuation of the solution from the four-sphere
to the domain of real $s$ tends to zero as  
Аналитическое $s$ tends to infinity and 
$\psi $ tends to zero as $\psi \sim e^{-2s}$.

To construct the solution on the sphere, 
let us introduce new angular variables
$\sigma ',\psi '$
that are related to $\sigma ,\psi $ as follows,
\begin{equation}
       \cos{\sigma '} = \sin{\sigma }\cos{\psi },~~~
       \sin{\sigma '}\cos{\psi '} = \cos{\sigma }
\label{4.10}
\end{equation}
This change of variables corresponds to the shift of the north pole 
to the point 
$(\sigma =\frac{\pi }{2}, \psi =0)$. 
In terms of the new variables, the solution has to be symmetric with
respect to the meridian line
$\psi '=\frac{\pi }{2}$. Let us assume, following Coleman and De
Luccia, that 
$\phi $ has greater symmetry, namely, that it is $O(4)$-symmetric.
Then $\phi $ depends on one variable
$\sigma '$ only.  
Making use of this assumption, one simplifies the expresion for the action,
\begin{equation}
       S_\zeta = 2\pi^2 \int\limits_{0}^{\pi }d\sigma ' \sin^3\sigma'
       \left(\frac{1}{2H^2}\left(\frac{d\phi }{d\sigma '}\right)^2 +
       \frac{V(\phi )}{H^4}\right)
\label{4.11}
\end{equation}
and obtains the equation for $\phi $,
\begin{equation}
       \frac{d^2\phi }{d\sigma'^2} + \frac{3}{\tg{\sigma'}}
       \frac{d\phi}{d\sigma'} - \frac{V'(\phi )}{H^2} =0
\label{4.12}
\end{equation}
The conditions that are imposed on $\phi $ are:

a)
\begin{equation}
       \left.\frac{d\phi }{d\sigma '}\right|_{\sigma '=0} =
       \left.\frac{d\phi }{d\sigma '}\right|_{\sigma '=\pi } = 0
\label{4.13}
\end{equation}
(this follows from the requirement of smoothness of $\phi $);

b) $\phi $ is not constant on the sphere
(this is needed to satisfy C2, see below);

c) $\phi (\sigma '=0)$ is close to the false vacuum, i.e., to
$\phi=0$; 
$\phi (\sigma '=\pi )$ is close to the true vacuum, i.e., to
$\phi _+$.

Solutions with these properties exist for wide class of the scalar
potentials 
$V(\phi )$ which obey
(see refs.  \cite{CdL,HawkMoss,GL})
\begin{equation}
       \left|\frac{V''(\phi _0)}{H^2}\right| > 4
\label{4.14}
\end{equation}
On the other hand, if this inequality does not hold, the condition C2
cannot be satisfied
\cite{HawkMoss,GL}. The latter case will be considered in the next
Section, and here we assume that the non-trivial solution does exist.
The solution under discussion obeys C1 automatically due to the
$O(4)$-symmetry. We postpone the discussion of C2, and study now
the continuation of
$\phi $ 
into the region of real 
$\eta$.

\begin{figure}[htb]
        \begin{center}
        \epsfig{file=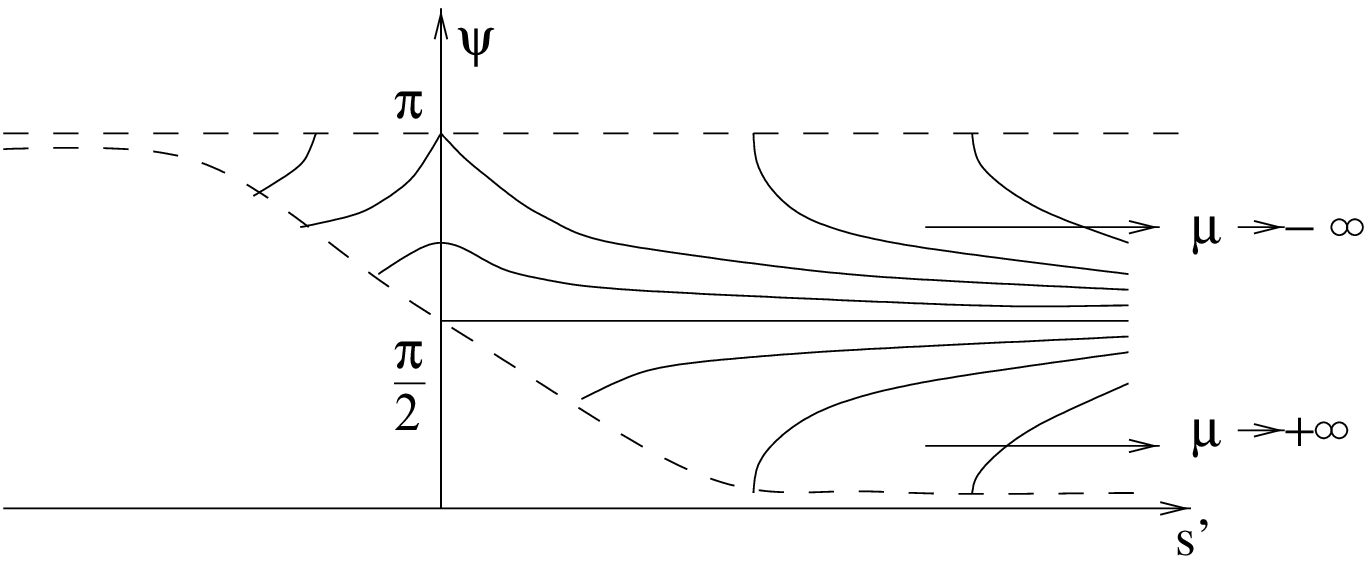, width=10cm}
        \caption{}
	\label{f7}
        \end{center}
\end{figure}

\begin{figure}[htb]
        \begin{center}
        \epsfig{file=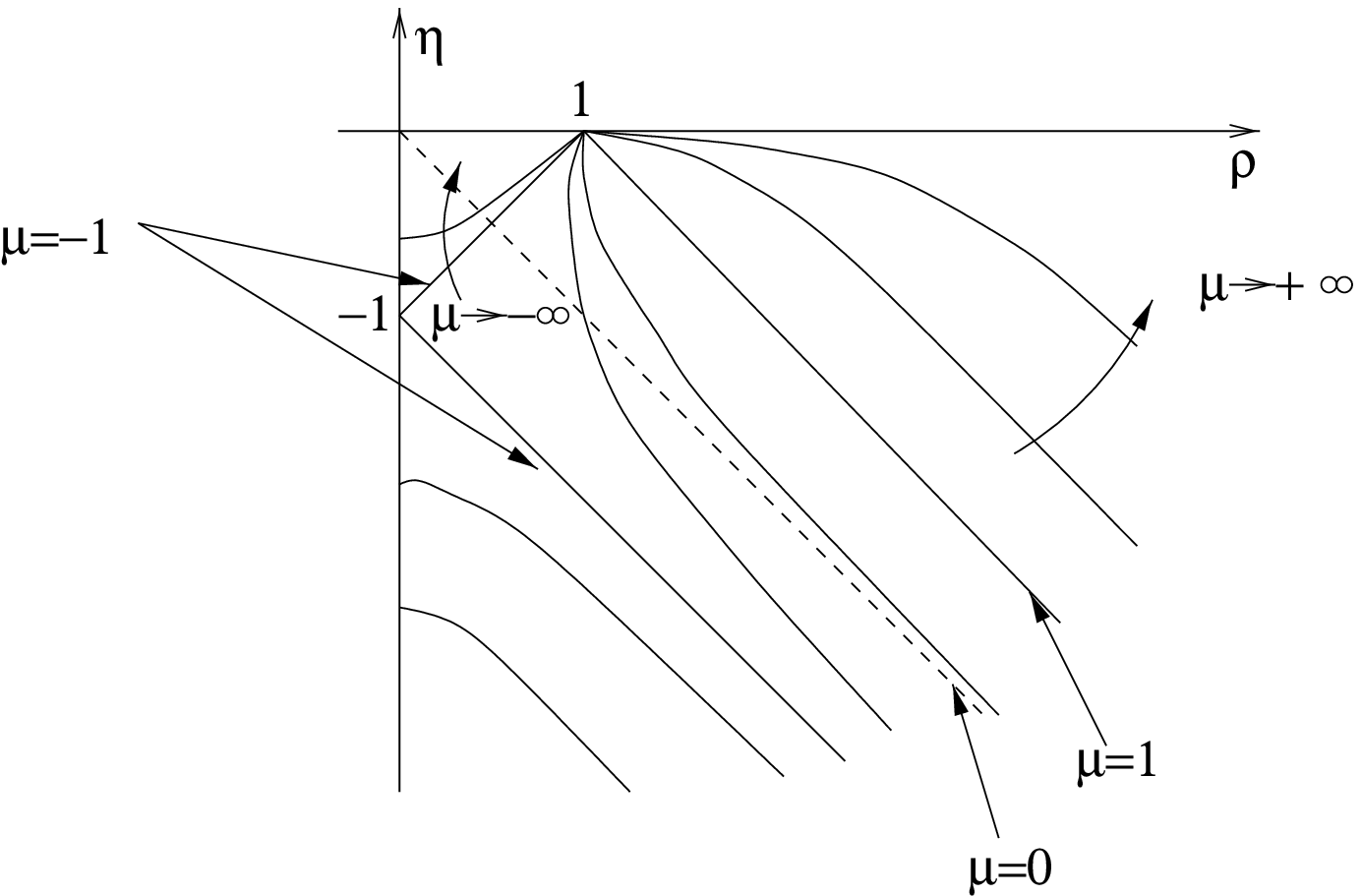, width=10cm}
        \caption{}
	\label{f8}
        \end{center}
\end{figure}

At this stage, it is convenient to regard 
$\phi $ as a function of the variable $\mu =\cos{\sigma '}$
rather than of $\sigma '$ itself.
According to eq. (\ref{4.10}), $\mu$ is expressed in terms of the
coordinates introduced in Section 4,
\begin{equation}
       \mu =\sin{\sigma }\cos{\psi }=i\sh{s}\cos{\psi }=\ch{s'}\cos{\psi }
\label{4.15}
\end{equation}
We show in fig. \ref{f7} the lines of constant 
$\mu $ in cooordinates $(s',\psi )$
(we display the region $\cos{\psi } < \th {s}$ that corresponds to the
physical de Sitter space)
By comparing figs. \ref{f7} and \ref{5f}, 
we obtain the set of lines of constant $\mu$,
and hence $\phi $, on the plane $(\eta,
\rho )$, see fig.  \ref{f8}.
It is clear from fig. \ref{f8} that $\lim_{\eta\to 0^-,\rho >1} \phi
= \lim_{\mu \to +\infty} \phi $, and $\lim_{\eta\to 0^-,\rho <1} \phi
= \lim_{\mu \to +\infty} \phi $.  
So, to understand the behavior of $\phi$, one has to consider its
asymptotics at
$\mu
\to\pm\infty$.

As $\mu $  tends to $+\infty$, the variable $\sigma'$ behaves as follows,
$\sigma '=i\sigma ''$, $\sigma ''$ 
is real and tends to $+\infty$.
In the regime $\mu \to -\infty~$,  $\sigma '$ equals $\pi +i\sigma ''$
where $\sigma ''$
is again real and tends to $+\infty$. The field equations in these
regions are obtained by analytically continuing eq. 
(\ref{4.12}):
\begin{equation}
       \frac{d^2\phi }{d\sigma''^2} + \frac{3}{\th{\sigma ''}}
       \frac{d\phi }{d\sigma ''} + \frac{V'(\phi )}{H^2} =0
\label{4.16}
\end{equation}
The initial conditions that supplement eq. (\ref{4.16}) in the two
cases
are as
follows:

$\left.\phi \right|_{\sigma''=0}$ is close to $0$ as  $\mu \to
+\infty$

$\left.\phi \right|_{\sigma''= 0 }$ is close to $\phi _+$ as
$\mu \to -\infty$

$\left.\frac{d\phi}{d\sigma''}\right|_{\sigma''=0} =0$ 
in both cases.

\begin{figure}[htb]
        \begin{center}
        \epsfig{file=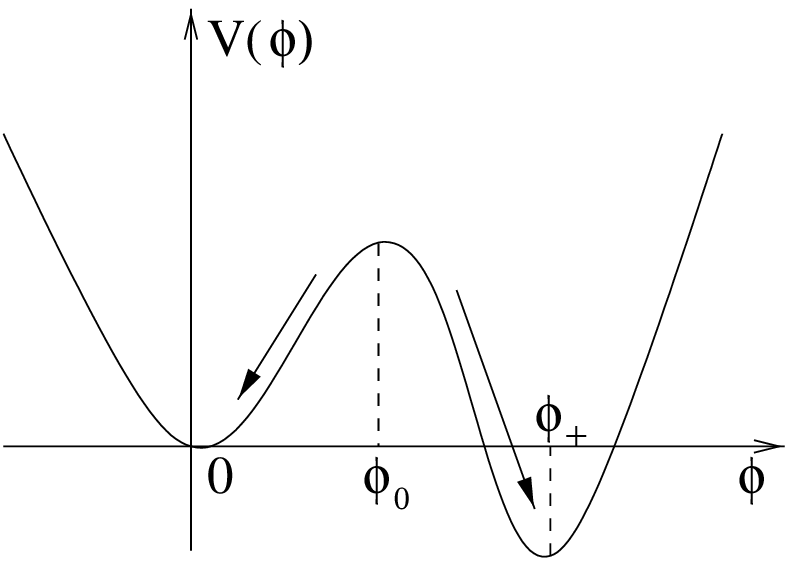, width=5cm}
        \caption{}
	\label{f9}
        \end{center}
\end{figure}

 Equation (\ref{4.16}) may be viewed as an equation for a particle with
 coordinate
 $\phi $ 
in time $\sigma ''$. This particle moves classically in the
 potential
$\frac{V(\phi )}{H^2}$ with time-dependent friction coefficient
$\frac{3}{th \sigma''}$. Since the friction coefficient is larger that
 3 at all times, the particle moving with zero initial velocity will
 always reach the closest minimum of the potential, provided that its
 initial position is different from
$\phi _0$. In the former and latter cases this will be 
$\phi =0$ and $\phi  =\phi _+$, respectively, see fig.
\ref{f9}.

Therefore, $\phi \to0$ as $\mu \to+\infty$, while $\phi \to\phi _+$
as $\mu \to-\infty$.  This means that
\begin{equation}
       \left.\phi \right|_{\eta \to 0^-, \rho >1} =0,~~~
       \left.\phi \right|_{\eta \to 0^-, \rho <1} =\phi _+,
\label{4.18}
\end{equation}
It follows from the latter relations  that the solution describes a
bubble of the true vacuum in the false one, whose coordinate size 
is asymptotically equal to
$\rho = 1$. The bubbles of other sizes are obtained from this solution
by scaling of coordinates, i.e., by the change
$\eta
\to \alpha\eta, \rho \to \alpha\rho $; the leading
semiclassical exponent of their formation is independent of their
size as the action does not change under scale transformations.

\begin{figure}[htb]
        \begin{center}
        \epsfig{file=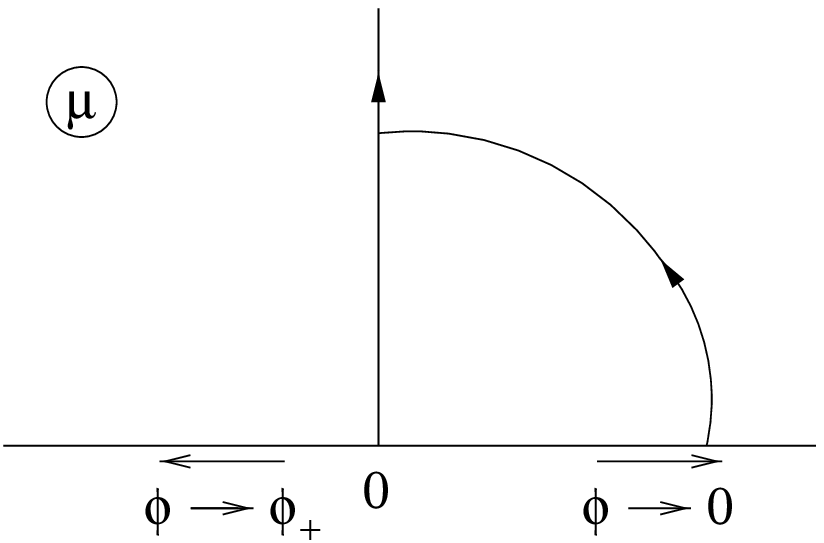, width=5cm}
        \caption{}
	\label{f11}
        \end{center}
\end{figure}

Let us come back to the condition C2.  
The behavior of the coordinates
$s$ and $\psi $ that enters this condition
corresponds, according to eq. (\ref{4.15}), to the following
behavior of the variable
$\mu $,
\begin{equation}
       \mu \sim ie^s\cos{e^{-2s}} \sim ie^s,~~s\to\infty
\end{equation}
What remains to be done is to analytically continue $\phi $ 
into the region of large
$\mu $. Some care must be taken when performing this continuation,
though. It is convenient to recall that the continuation from the
region of real $\eta$ to purely imaginary $\eta$ (and, hence, real
$\zeta$) is to be done as shown in fig.
\ref{f3}.
This means that $s$ changes from
$s_1-i\frac{\pi }{2}$ to
$s_1$, where $s_1$ 
is real and large
(recall the considerations that have lead to eq.
(\ref{4.5})). The variable $\mu $ remains in the first quadrant, being
far away from the origin, see fig. \ref{f11}.
With this continuation one has $\lim_{\mu \to +i\infty} \phi = \lim_{\mu \to
+\infty} \phi =0$, so the condition C2 is indeed satisfied.

We conclude that the Coleman--De Luccia instanton indeed describes the
false vacuum decay in the de Sitter space-time with the qualification that
the quantum fluctuations
of the scalar field are initially in the state of conformal vacuum.

\section{ Hawking--Moss instanton as the limit of 
 constrained instantons} 

Let us now consider the case when a non-trivial solution of
the field equation on the four-sphere does not exist. To obtain the
false vacuum decay rate, we have to modify our previous arguments.
The main idea is that instead of true solution to the field equation,
one considers a family of almost saddle point configurations
$\phi_{\varepsilon}({\bf x},\eta)$
which obey the boundary conditions of Section 3 and contribute to the
path integral.
The action on these configurations depends on the parameter
$\varepsilon$, and their contribution to the decay rate tends to
maximum as
$\varepsilon \to 0^{+}$. The limiting configuration
$\phi_{\varepsilon = 0}$ (Hawking--Moss instanton) does not obey the
boundary conditions (even though it is a solution to the field
equation);
however, its action determines the semiclassical exponent of the
decay.
The situation in this case resembles one that emerges in the study of
instanton effects in four-dimensional Yang--Mills--Higgs theories
\cite{Affleck}.

Let us introduce ``unity'' in the original path integral (\ref{2.3}),
\[
      1= \int\limits_{-\infty}^{\infty} d\varepsilon~
	\delta(\varepsilon-\Delta) =\int\limits_{-\infty}^{\infty}
	d\varepsilon \int\limits_{-\infty}^{\infty}
	\frac{d\alpha}{2\pi}~  
	e^{i\alpha(\varepsilon -\Delta)}
\]
where
\[
	\Delta =\int \frac{d^3xd\eta}{2H^2\eta^2} 
	\left(\left(\frac{\partial\phi_1}{\partial\eta}\right)^2 +
	\left(\frac{\partial\phi_2}{\partial\eta}\right)^2 -
	\left(\frac{\partial\phi_1}{\partial x^i}\right)^2 -
	\left(\frac{\partial\phi_2}{\partial x^i}\right)^2\right)
\]
we now change the order of integration, and integrate over 
$d\varepsilon d\alpha$
in the very end.
After repeating, with slight modifications, the arguments of Sections
3 and 5, we obtain for the decay rate
\begin{equation}
       \Gamma =\int d\varepsilon \exp{(-S_\zeta [\phi _\varepsilon ])}
\label{5.2}
\end{equation}
where $S_\zeta $ is still defined by eq. (\ref{4.11}), while
the classical solutions $\phi _\varepsilon $ obey, besides all
conditions of Section 5, an extra constraint,
\begin{equation}
        \frac{2\pi ^2}{H^2} \int\limits_{0}^{\pi } d\sigma '
        \sin^3\sigma' ~\frac{1}{2}
        \left(\frac{\partial\phi _\varepsilon }{\partial\sigma '}\right)^2
        =\varepsilon
\label{5.3}
\end{equation}
To obtain the equation for  $\phi _\varepsilon $ 
one subtracts from the action (\ref{4.11}) the constraint
(\ref{5.3}) with the Lagrange multiplier $\lambda $,
\begin{eqnarray}
      S_\zeta &-& \lambda \frac{2\pi ^2}{H^2} \int\limits_{0}^{\pi }  
	d\sigma'
        \sin^3\sigma' ~\frac{1}{2}
        \left(\frac{\partial\phi _\varepsilon }{\partial\sigma
        '}\right)^2 \nonumber\\ 
	&=&
          \frac{2\pi ^2}{H^2} (1-\lambda ) \int\limits_{0}^{\pi }
        d\sigma ' \sin^3\sigma' \left(\frac{1}{2}
        \left(\frac{\partial\phi _\varepsilon }{\partial\sigma
        '}\right)^2 + \frac{V(\phi _\varepsilon )}{H^2(1-\lambda )}\right)
\label{5.4}
\end{eqnarray}
and varies this expression
with respect to $\phi _\varepsilon $. The resulting equation is
\begin{equation}
       \frac{d^2\phi_\varepsilon }{d\sigma '^2} + \frac{3}{\tg{\sigma '}}
       \frac{d\phi_\varepsilon }{d\sigma '} -
       \frac{V'(\phi _\varepsilon)}{H^2(1-\lambda )} =0
\label{5.5}
\end{equation}
We find that, as far as the field equation is concerned, 
the effect of the constraint (\ref{5.3}) is the
modification of the scalar potential.

The boundary problem consisting of equation
(\ref{5.5}), the constraint (\ref{5.3}) and conditions (a), (b) and
(c) of section 5 has a solution at every positive and
 sufficiently small 
$\varepsilon$. To see this, we note that the solution obeying eq.
(\ref{5.5})  and conditions (a), (b), (c) exists, provided that
\[
      \left|\frac{V''(\phi _0)}{H^2(1-\lambda )}\right| >4
\]
(cf. eq. (\ref{4.14})). At
\[
       (1-\lambda) \to \left|\frac{V''(\phi _0)}{4 H^2}\right|
\]
this solution tends to a constant on the four-sphere; the integral of
its gradient squared tends to zero, so the constraint (\ref{5.3}) is 
indeed
satisfied with an appropriate choice of 
$\lambda$.

From the results of Section 5 it follows that at a given
$\varepsilon >0$, the analytical continuation of the solution
$\phi _\varepsilon $  to physical space-time describes
the bubble of the true vacuum. On the other hand, as
$\varepsilon \to 0^+ $, the configuration
$ \phi _\varepsilon $   tends to a constant field on the four-sphere,
$\phi = \phi _0$, i.e., it becomes the Hawking--Moss instanton.
It is known \cite{HawkMoss,GL}
that in the models under discussion, this instanton is an absolute
minimum of the action
$S_\zeta$, at least among the $O(4)$-symmetric configurations.
Hence,
$S_\zeta [\phi _\varepsilon ]$ as a function of $\varepsilon $ has a
minimum at $\varepsilon =0$, 
so the largest contributions into the rate
(\ref{5.2}) come from the solutions $\phi _\varepsilon $ at small
$\varepsilon$. 
The integral (\ref{5.2}) is then determined by the limiting point
$\varepsilon = 0$ and is equal to
\[
       \Gamma \propto e^{-S_\zeta [\phi _0]}
\]
where the exponent is the action of the Hawking--Moss instanton,
\[
       S_\zeta [\phi _0] =\frac{8\pi ^2}{3H^4} V(\phi_0)
\]

Thus, the Hawking--Moss instanton is the limiting case of the
constrained instantons,
$\phi_{\varepsilon}({\bf x}, \eta)$.
The configuration $\phi_{\varepsilon}({\bf x}, \eta)$ at
$\varepsilon > 0$ has the same structure as the Coleman--De Luccia
instanton considered in the previous Section. We conclude that the
false vacuum decay again occurs through the formation of bubbles of
the new phase.

\section{Jumps up}

Finally, let us discuss an interesting property of the scalar field in
the 
de Sitter space-time which is absent in the case of flat space. In the
de Sitter space-time, there is a possibility of tunneling
transitions from true vacuum to false one; the possibility of ``jumps
up'' due to quantum fluctuations of the scalar field was pointed out
in ref.  
\cite{Linde83} in the context of inflationary theories.
The energy needed for these transitions comes from the gravitational
field; within the approximation made in this paper, the latter
represents, in fact, an infinite source of energy. The description of
these transitions is natural in the framework advocated in this paper.

Indeed, when introducing the coordinates $\sigma ',\psi '$ (see
Section 5) we could write, instead of eq. 
(\ref{4.10}), another change of variables, 
\[
       \cos{\sigma '} = \sin{\sigma }\cos{(\psi -\psi _0)},~~~
       \sin{\sigma '}\cos{\psi '} = \cos{\sigma }
\]
where $\psi _0$ is a parameter between $0$ and
$\pi $. 
This would correspond to the shift of the pole on the four-sphere to
the point
$\sigma =\frac{\pi}{2}$, $\psi =\psi _0$.  
There again exists an $O(4)$-symmetric solution of eq. (\ref{4.8}) 
in this coordinate system, that depends on 
$\sigma '$ only and coincides, as the function of $\sigma '$, 
with the
solution considered in Section 5.
(We discuss here the case when the non-trivial solution of Coleman--De
Luccia type exists; the considerations that follow are
straightforwardly generalized to the Hawking--Moss case.)
However, this solution does not make sense at arbitrary
$\psi_0$.
This is clear from the fact that its analytical continuation to real
$\rho $ and $\eta $ does not obey the smoothness property
$\left.\frac{\partial\phi
}{\partial\rho} \right|_{\rho =0} =0$.  Indeed, let us introduce, as
in Section 5, the variable
$\mu = \cos{\sigma'}$, 
which is now related to $s'$
and $\psi$  by
\[
       \mu =\ch{s'}\cos{(\psi -\psi _0)}
\]
(cf. eq. (\ref{4.15})). Making use of eq. (\ref{3.8}), we obtain
\[
       \left.\frac{\partial\phi }{\partial\rho }\right|_{\rho =0}
       = \frac{e^{s'}}{\ch{s'}} \left.\frac{\partial\phi }{\partial\psi }
       \right|_{\psi =\pi } = -\ch{s'}\sin{(\pi -\psi _0)}
       \left.\frac{d\phi }{d\mu }\right|_{\mu = ch s'\cos{(\pi -\psi _0)}}
\]
which is not equal to zero if $\psi _0 \ne 0,\pi $. 
Hence, the solution can describe a physical process only if
$\psi _0$ equals  $0$ or $\pi $. 
The former possibility has been considered in Section 5. Let us now
turn to the case $\psi _0 = \pi $.
Instead of eq. (\ref{4.15}) we have in this case
\[
        \mu = -i\sh{s}\cos{\psi } = -\ch{s'}\cos{\psi }
\]
Hence (cf. eq. 
(\ref{4.18}))
\[
       \lim_{\eta \to 0^-, \rho >1} \phi =\lim_{\mu \to -\infty} \phi
       =\phi_+,~~~
       \lim_{\eta \to 0^-, \rho <1} \phi =\lim_{\mu \to +\infty} \phi
       =0,~~~
       \lim_{s\to\infty, \phi \sim e^{-2s}} \phi =\phi_+
\]
These are precisely the properties that one would anticipate for the
transitions from the true vacuum to the false one. This means that the
solution with
$\psi_0 = \pi$ describes the formation of a bubble of the false vacuum
in the true one. The probability of this process is again determined
by the action
(\ref{4.11}), but now one has to subtract the action of the trivial
solution corresponding to the true vacuum,
\[
       \frac{2\pi ^2}{H^4} \int\limits_{0}^{\pi }d\sigma '\sin^3\sigma'~
       V(\phi _+) = \frac{8\pi ^2}{3H^4} V(\phi _+)
\]
In this way we obtain the probability of the bubble formation,
\[
  \Gamma \propto\exp{\left[ -S_\zeta +\frac{8\pi ^2}{3H^4} V(\phi_+)
                       \right]}
\]
The limiting case of flat space-time is obtained by sending the
Hubble parameter to zero. 
The action $S_\zeta $ tends to a constant in this limit, this constant
being the exponent of the false vacuum decay in flat space.
Since $V(\phi _+) <0$, the probability of the transition from the true
vacuum to the false one vanishes in this limit, as expected.

\section{Conclusions}

In this paper, the false vacuum decay in the de Sitter space-time was
discussed within the path integral approach. We considered the
theory of one scalar field and chose conformal vacuum of the
scalar particles above the false classical vacuum as a definition of
the quantum false vacuum. In this case we reproduced from the first
principles the Coleman--De Luccia and Hawking--Moss results for the
leading semiclassical exponents of the decay rate.

We gave the interpretation of the Hawking--Moss instanton as a
limiting case of constrained instantons and observed that in spite of
the fact that it determines the transition probability, the structure
of the field after tunneling is
in fact determined by other configurations that are close to this
instanton on the four-sphere. The result of the tunneling process is,
as usual, the bubble of the true vacuum.

We discussed yet another interesting phenomenon, namely, the
transitions from the true vacuum to the false one, and calculated the
probability of this process. This probability tends to zero in the
limit of flat space-time, in accord with expectations.

Though we concentrated in this paper on a particular class of
processes, our approach that explicitly accounts for the initial state
of the quantum field (in our case it was the conformal vacuum) is
straightforward to generalize to other semiclassical processes in
curved space-time. The key ingredients of our approach --- the
boundary conditions
(\ref{2.15b}) and (\ref{2.17}) --- are in fact quite general,
provided the positive-frequency conditions are understood in terms of
asymptotic properties of the solution
$\phi(x)$ with respect to the set of modes appropriate to a given
vacuum of the quantum field. It is another matter that the concrete
analytical continuations of the solution into the domain of complex
coordinates, which were used in this paper, are likely to be
specific to the de Sitter space and to the conformal vacuum of scalar
particles. 

As a final remark, let us point out that there are choices of
coordinates in the de Sitter space-time that differ from eq.
(\ref{1.2}) and, correspondingly, the choices of the vacuum of the
scalar particles that differ from the conformal vacuum.
In particular, Derulle
\cite{Deruel} considered the false vacuum decay, in the 
background metrics approximation, 
making use of the static coordinate system, and
reproduced the Coleman--De Luccia and Hawking--Moss results as well.
On the other hand, the vacuum defined with respect to the static
coordinate system is not invariant even under translations
\cite{Birel}
and, therefore, it is quite different from the vacuum considered in
this paper. So, the coincidence of the results of our paper and ref.
\cite{Deruel} appears surprizing.
It is worth noting, however, that the quantum vacuum was not
explicitly defined in ref.
\cite{Deruel}.
We think that this coincidence may be explained as follows: in effect,
the analysis of ref.
\cite{Deruel} applies not to tunneling from the vacuum of the static
coordinate system, but to transitions from 
the conformal vacuum which looks
as a highly excited state from the point of view of
the static coordinate system.
This point deserves further analysis.

We hope that this paper clarifies certain aspects of the path integral
fomalism  in curved space-time and 
opens up new ways of its application to
various semiclassical processes.

\end{document}